\documentstyle[12pt,epsf]{article}

\newcommand\beq{\begin{equation}}
\newcommand\eeq{\end{equation}}
\newcommand\beqa{\begin{eqnarray}}
\newcommand\eeqa{\end{eqnarray}}

\newcommand\bsigma{\mbox{\boldmath$\sigma$}}
\newcommand\btau{\mbox{\boldmath$\tau$}}
\newcommand\bpi{\mbox{\boldmath$\pi$}}
\newcommand\btheta{\mbox{\boldmath$\theta$}}
\newcommand\balpha{\mbox{\boldmath$\alpha$}}

\newcommand\bk{{\bf k}}
\newcommand\br{{\bf r}}

\title{Pseudoscalar Mesons in Nuclear Medium
\footnote{\uppercase{T}his work is supported by the Grant-in-Aid for 
Scientific Research Fund of the Ministry of Education, Culture, Sports, 
Science and Technology (11640272, 13640282).}}

\author{T.~Tatsumi}

\date{}
\begin{document}

\maketitle

\vskip -0.8cm
\centerline{\it Department of Physics, Kyoto University, 
Kyoto, 606-8502, Japan}

\centerline{\it E-mail: tatsumi@ruby.scphys.kyoto-u.ac.jp}  


\begin{abstract}
The behavior of pseudoscalar mesons in nuclear medium is
reviewed with an emphasis on the possibility of their Bose-Einstein
condensation in dense matter. In particular pion condensation is
reexamined in detail, stimulated by recent theoretical and observational
developments.  
\end{abstract}

\section{Introduction}

%hadron-quark deconfinement

Recently much attention has been paid for high density QCD: hadron matter 
at relatively low density and quark matter at high density are typical 
subjects there. As a key concept which goes through hadron and quark worlds 
chiral symmetry is realized in both matter, but in a different way.
Parity-even and odd quantities interplay in this context. Since the vacuum is 
good parity state, there is no expectation value of parity-odd 
operators. However, they may have finite values in matter (parity violation); 
Bose-Einstein condensation of pseudoscalar mesons in hadronic matter or 
nonvanishing of the parity-odd mean-field, 
$\langle\bar q\gamma_5\Gamma^\alpha q\rangle$, in quark matter. 
We consider the 
particle-hole operator here and discuss how the pseudoscalar quantity
becomes nonvanishing in hadronic matter and what are its implications, 
by studying 
the behavior of pseudoscalar mesons in nuclear medium.  

Pseudoscalar mesons ($\pi, K$) have some salient features; they are the 
lightest hadrons without and with strangeness and considered as the Nambu-
Goldstone bosons as results of spontaneous breaking down of chiral symmetry.
Since they are bosons, they may lead to the Bose-Einstein condensation in 
some situations. 

To understand the behavior of these mesons in nuclear 
medium, we also take into account the effects of resonances. $\Delta(1232)$ 
strongly couples with nucleon by the $p$- wave $\pi N$ interaction and 
$\Lambda(1405)$ gives rise to a peculiar feature in $s$- wave 
$KN$ scattering near threshold. The mass difference between these resonances 
and nucleons is small ($O(m_\pi)$), so that they should play important 
roles even for low energy phenomena (energy-momentum scale 
$\sim O((2 - 3)m_\pi)$) we are interested in. 
Thus chiral symmetry and resonances 
may characterize the behavior of these mesons in nuclear medium.  

\begin{figure}[ht]
%\epsfxsize=10cm   %width of figure - will enlarge/reduce the figures
%\epsfbox{fig3.eps}
%\figurebox{2cm}{3cm}{} %to have a box alone 
\centerline{\epsfxsize=2.5in\epsfbox{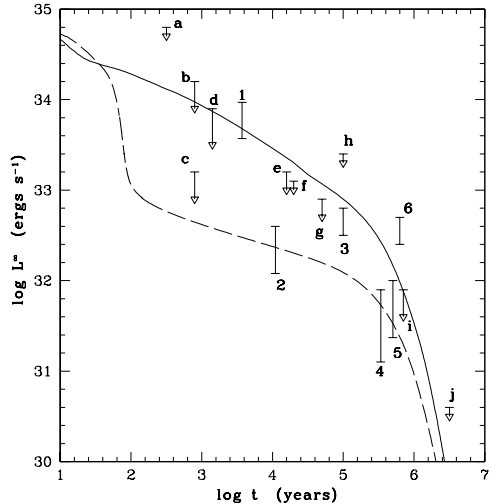}}   
\caption{Cooing curves for stars with medium FP EOS, taken from 
ref.2. Solid curve shows standard cooling of a 1.2$M_\odot$ neutron 
star and dashed curve shows pion cooling of a 1.4$M_\odot$ star. 
The data of 
$c$,2,4, 5 indicate cooler stars: (c)3C58, (2) Vela, (4) Geminga and (5)
RXJ 1856-3754.}
\end{figure}

There have been performed and planed many nuclear experiments to reveal the 
behavior of these mesons in nuclear medium. On the other hand, observations 
of compact stars have also provided information about it. Very recently 
appeared an interesting data about the surface temperature of young 
pulsar \cite{sla}. They 
reported that the pulsar inside the historical supernova 3C58 shows 
too low surface temperature to be explained by the standard cooling 
scenario, which 
essentially assumes usual neutron matter inside it (see Fig~1). The 
importance of 
this observation is in the age ($\sim 10^3$yr) of this pulsar. As we shall 
see later if this fast cooling is attributed to the presence of pion 
condensation, the star cools very rapidly in the early neutrino-emitting 
phase and the difference 
of the surface temperature from the standard cooling scenario 
becomes remarkable there. So we can hope to see 
the evidence of pion condensation more clearly for young pulsars. 
In Fig.~1 we present our theoretical cooling curve with observational
data. We can see that pion cooling can explain cooler stars including 
3C58 \cite{tsu}.

\section{Kaon Condensation}

The low-energy $KN$ interaction is specified by three kinds of the $s$- wave 
interaction, 
the $KN$ $\sigma$ term, the energy-dependent Tomozawa-Weinberg term and the 
resonant term with $\Lambda(1405)$. Empirically the scattering amplitude 
shows an interesting behavior in the $I=0$ channel due to the existence 
of the resonance $\Lambda(1405)$ below the threshold. Recently some people 
propose a possibility of the deeply bound kaonic nuclei in relation to 
the property of this resonance \cite{kis}.
On the other hand, the interaction becomes relatively weak in the $I=1$ 
channel, and it is relevant when we consider kaon ($K^-$)) condensation 
in neutron stars.

Since we know the non-resonant $s$- wave terms can be well described in 
terms of 
$SU(3)\times SU(3)$ chiral symmetry, kaon condensation has been discussed 
on the basis of chiral Lagrangians \cite{mut,ty}.  These $s$- wave terms 
cooperatively
work to give a large decrease of the effective energy of kaons in nuclear
medium. When the energy reaches the electron chemical potential as
density increases, kaons
begin to condense through the reaction,
\beq
nn\rightarrow npK^- . 
\eeq  
This is very similar to the Bose-Einstein condensation of alkali atoms 
\cite{mut}.

The most important consequence of kaon condensation is the large
softening of the equation of state, which leads to an interesting
phenomenon, {\it delayed collapse} of protoneutron stars 
to produce the low-mass black holes \cite{bro,bau,yt,pon,pon2}.

%delayed collapse

\section{Pions in Nuclear Medium}

%Landau Fermi liquid theory 

Since pions couple with particle -hole and $\Delta$- hole states with
the same quantum number, we can study the properties of these states as 
well as pions themselves by considering the pion propagation in nuclear 
medium \cite{mig}.
In Fig.~2 we show the longitudinal spin-isospin excitation (pionic) modes 
in the energy ($\omega$)-momentum ($k$) plane.
\begin{figure}[ht]
%\epsfxsize=10cm   %width of figure - will enlarge/reduce the figures
%\epsfbox{fig3.eps}
%\figurebox{2cm}{3cm}{} %to have a box alone 
\centerline{\epsfxsize=3in\epsfbox{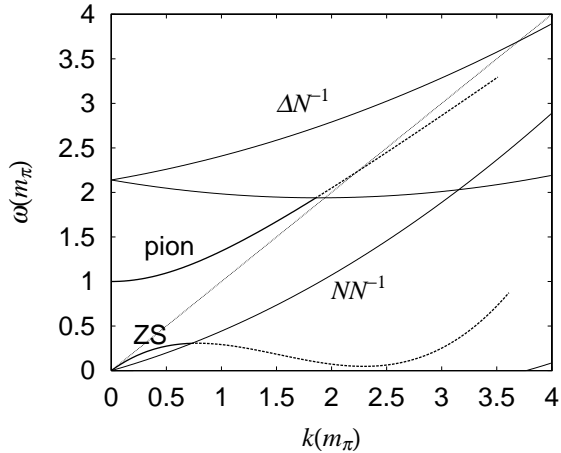}}   
\caption{Schematic view of the spin-isospin modes in symmetric (N=Z) nuclear 
matter. $NN^{-1}$ and $\Delta N^{-1}$ denote the continuum spectrum of the 
particle-hole and $\Delta$-hole excitations, respectively. ``ZS'' means the 
zero-sound mode, which corresponds to the Gamow-Teller state in asymmetric 
matter.}
\end{figure}

\subsection{Pionic Excitations within the Fermi Liquid Theory}

Consider the pion propagator in nuclear medium at density $\rho_B$:
\beq
D^{-1}_\pi(\omega, {\bf k};\rho_B)=\omega^2-m_\pi^2-{\bf k}^2
-\Pi({\bf k},\omega; \rho_B),
\eeq
due to the self-energy $\Pi$. For the $p$- wave $\pi N$ interaction
\footnote{The $s$- wave interaction is negligible in symmetric ($N=Z$)
nuclear matter.}
,
it is simply given by the particle-hole polarization function $U^{(0)}$,
\beq
\Pi_p^{(0)}=-k^2\left(U_N^{(0)}(\omega, {\bf k}; \rho_B)
+U_\Delta^{(0)}(\omega, {\bf k}; \rho_B)\right),
\eeq
in the lowest order, where we have taken into account the 
nucleon particle - hole (ph) and 
$\Delta$- hole ($\Delta$h) states
\footnote{We, hereafter, use the nonrelativistic approximation for
nucleons. See ref.~12 for a relativistic treatment.}
. The 
polarization functions $U^{(0)}_\alpha$ are 
further given in terms of the Lindhard functions $L_\alpha$ \cite{fet},
\beq
U^{(0)}_\alpha=\left(\frac{f_{\pi N\alpha}\Gamma(k)}{m_\pi}\right)^2
L_\alpha,
\eeq
where we introduced the form factor 
$\Gamma=(\Lambda^2-m_\pi^2)/(\Lambda^2+k^2)$ with 
the cut-off $\Lambda\sim O(1{\rm GeV})$, and the $\pi NN$ and 
$\pi N\Delta$ coupling constants are $f_{\pi NN}\sim 1$ and 
$f_{\pi N\Delta}\sim 2 ({\rm Chew-Low~ value})$, respectively.
the Lindhard functions $L_\alpha$ are explicitly evaluated to be, e.g. 
\beq
L_N={\rm Re}L_N+i{\rm Im}L_N, 
\eeq
with
\beqa
{\rm Re}L_N&=&\frac{2m_N^*}{\pi^2}(p_F^p\phi_p(k,\omega)+
p_F^n\phi_n(k,-\omega)),\nonumber\\
\phi_i(k,\omega)&=&\frac{m_N^{*2}}{2k^3p_F^i}\left\{-ab^i+
\frac{a^2-b^{i2}}{2}\ln\left|\frac{a+b^i}{a-b^i}\right|\right\}\nonumber\\
a&=&\omega-k^2/2m_N^*,~b^i=kv_F^i,
\eeqa
for $\pi^-$ propagation. Here we introduced the effective mass $m_N^*$
for nucleons.

\begin{figure}[ht]
%\epsfxsize=10cm   %width of figure - will enlarge/reduce the figures
%\epsfbox{fig3.eps}
%\figurebox{2cm}{3cm}{} %to have a box alone 
\centerline{\epsfxsize=2.5in\epsfbox{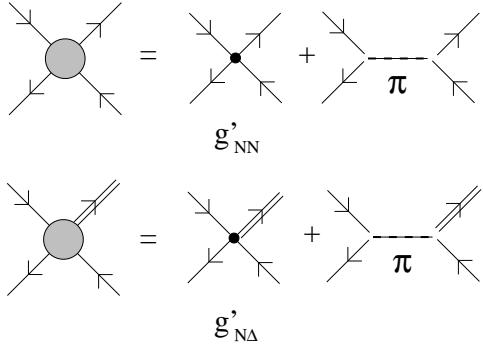}}   
\caption{Examples of the particle-hole and $\Delta$- hole 
interactions in the 
spin-isospin channel. They are written as sum of the one-pion exchange 
interaction and the phenomenological zero-range 
interaction with the Landau- Migdal parameters. }
\end{figure}

It is well known that the lowest order calculation is not sufficient 
to discuss the behavior of the pion in nuclear medium: we must take 
into account the correlations between ph and $\Delta$h states.
They can be easily incorporated in the spirit of Landau Fermi-liquid theory.  
Since we are interested in the region of 
$\omega, k\sim (2 -3 )m_\pi$,
full ph-ph, ph-$\Delta$h or $\Delta$h-$\Delta$h interaction should be 
separated into two terms, depending on their length scales: we  
explicitly treat the long-range ($O(m_\pi^{-1})$) interactions by way of  
pion, ph and $\Delta$h    
propagation, while the short-range ($O(m_N^{-1}\sim 0.2 {\rm fm})$) 
interactions are replaced by the 
momentum-independent local interactions parametrized by the Landau-Migdal 
parameters; e.g. 
\beq
{\mathcal F}_{NN}=f_{NN}+f'_{NN}\btau_1\cdot\btau_2+
{g_{NN}\bsigma_1\cdot\bsigma_2
+\left(\frac{f_{\pi NN}^2}{m_\pi^2}  \right)
g'_{NN}\bsigma_1\cdot\bsigma_2\btau_1\cdot\btau_2},
\eeq
where the strength in the spin-isospin channel is measured in the pion unit. 
It can be extended to include the isobar degrees of freedom: 
\beqa
{\mathcal F}_{N\Delta}
&=&\left(\frac{f_{\pi NN}f_{\pi N\Delta}}{m_\pi^2}\right)g'_{N\Delta}\bsigma_1\cdot
{\bf S}_2\btau_1\cdot{\bf T}_2\nonumber\\
{\mathcal F}_{\Delta\Delta}
&=&
\left(\frac{f_{\pi N\Delta}^2}{m_\pi^2}\right)g'_{\Delta\Delta}{\bf S}_1\cdot
{\bf S}_2{\bf T}_1\cdot{\bf T}_2
\eeqa
with the transition spin and isospin operators, ${\bf S}$ and ${\bf T}$.

Then we have the $p$- wave self-energy of the pion
 by considering the one-
line irreducible diagrams and the Dyson equations;
\beq
\Pi_p=\Pi_N+\Pi_\Delta,
\eeq
where
\beqa
\Pi_N&=&-k^2U_N=-k^2U_N^{(0)}[1+(g'_{\Delta\Delta}-g'_{N\Delta})
U_\Delta^{(0)}]/D,\nonumber\\
\Pi_{\Delta}&=&-k^2U_\Delta=-k^2U_\Delta^{(0)}
[1+(g'_{NN}-g'_{N\Delta})U_N^{(0)}]/D
\eeqa
with 
\beq
D=1+g'_{NN}U_N^{(0)}+g'_{\Delta\Delta}U_\Delta^{(0)}+(g'_{NN}g'_{\Delta\Delta}
-g'^{2}_{N\Delta})U_N^{(0)}U_\Delta^{(0)}.
\eeq

We can study the spin-isospin modes in another way, starting from the ph and 
$\Delta$h propagation within RPA. Both ways are equivalent with each 
other for the longitudinal modes. Considering the correlation function 
between the generalized spin-isospin density operator,
\beq
{\mathcal O}=\psi^\dagger\btau\bsigma\psi
+\frac{f_{\pi N\Delta}}{f_{\pi NN}}\psi_\Delta^\dagger{\bf T}{\bf S}\psi,
\eeq
we have the same excitation spectra of the spin-isospin modes as before 
\cite{shi,ich}.
Indeed   
the nuclear response function in the longitudinal spin-isospin 
channel is defined as follows;
\begin{eqnarray}
R(\omega, {\bf k})&=&\frac{{\rm Im}D_\pi(\omega, {\bf k};
\rho_B)}{D_0^2(\omega, {\bf k})}\nonumber\\
&=&\frac{1}{D_0^2(\omega, {\bf k})}\frac{{\rm Im}\Pi_p(\omega, {\bf k}; \rho_B)}
{[\omega^2-m_\pi^2\!\!-{\bf k}^2-{\rm Re}\Pi_p(\omega, {\bf k}; \rho_B)]^2
+[{\rm Im}\Pi_p(\omega, {\bf k}; \rho_B)]^2},
\end{eqnarray}
with the free pion propagator $D_0(\omega,{\bf k})=(\omega^2-m_\pi^2-{\bf
k}^2)^{-1}$.

\subsection{Pion Condensations}

\subsubsection{Neutral Pion Condensation}

First consider neutral pion condensation in symmetric (N=Z) nuclear matter, 
which require the following condition:
the softening of the longitudinal spin-isospin mode
\beq
R(0, k_c)\rightarrow \infty, ~~~{\rm Im}\Pi_p(0, k_c; \rho_c)\propto
\omega\theta(\omega) \rightarrow 0,
\eeq 
or equivalently
\beq
D^{-1}(0, k_c; \rho_c)\rightarrow 0,
\eeq
in terms of the pion propagator. It is to be noted that pion condensation 
by no means implies a naive Bose-Einstein condensation of pions, but the 
softening of the longitudinal spin-isospin mode with the critical 
momentum $k_c$. 
On the other hand, 
such mode is unstable and the Lindhard 
function has an imaginary part before 
the critical density $\rho_c$. We can see a peculiar enhancement of 
the strength 
function at small energy near the critical density.

The pion condensed phase can be represented in terms of chiral
transformation as follows:
\beqa
|\bpi^c\rangle&=&U(\btheta_V(\bk_c\cdot\br), \btheta_A(\bk_c\cdot\br))
|{\rm normal}\rangle\nonumber\\
&=&\exp\left[-i\left(\int\btheta_V\cdot{\bf V}^0d^3x
+{ \int\btheta_A\cdot{\bf A}^0d^3x}\right)\right]|{\rm normal}\rangle, 
\eeqa
with $\langle{\rm normal}|\bpi|{\rm normal}\rangle=0$.
Then
\beq
\langle\bpi\rangle=\bpi^c\neq 0.
\eeq
As an example,
\beq
{ \bpi^c=(0,0,A\cos k_cz)},~~~({ \pi^0}~{\rm  condensation}).
\eeq
It would be worth mentioning that $\pi^0$ condensation gives rise to a 
magnetic ordering of nuclear matter: it exhibits a liquid-crystalline nature 
with one-dimensional anti-ferromagnetic 
ordering, called Alternating-Layer-Spin [ALS] structure (see 
Fig.~4 ) \cite{sup}.  

\begin{figure}[ht]
%\epsfxsize=10cm   %width of figure - will enlarge/reduce the figures
%\epsfbox{fig3.eps}
%\figurebox{2cm}{3cm}{} %to have a box alone 
\centerline{\epsfxsize=2.5in\epsfbox{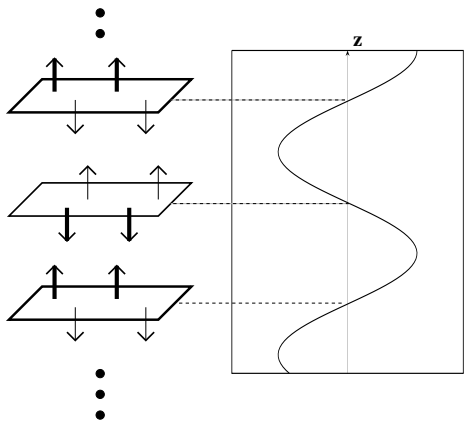}}   
\caption{Alternating-Layer-Spin [ALS] structure 
associated with $\pi^0$ condensate ($\propto A\cos k_cz$)
in symmetric nuclear matter. Bold arrows denote proton spins and thin arrows 
neutron spins.}
\end{figure}

Note that since {$U(\balpha)|{\rm normal}\rangle=|{\rm normal}\rangle$}, 
$U\in SU(2)$, in the isospin space, the isospin rotated condensate,
${\tilde{\bpi}}^c\!=\!{R}\bpi^c\!=\!(|\bpi^c|\sin\theta\cos\phi, 
|\bpi^c|\sin\theta\cos\phi,
|\bpi^c|\cos\theta), {R}\in O(3)$ 
is also possible with the same energy, which means {$\pi^\pm, \pi^0$} 
condensation. Indeed all the propagators for $\pi^0, \pi^\pm$ become 
identical in symmetric nuclear matter. Also note that 
$g'_{NN}$ should be replaced by $(g'_{NN}+g_{NN})/2<g'_{NN}$ in 
neutron ($Z=0$) matter.
Finally, 
since $\omega=0$ in this case, a potential description is 
possible instead of the  explicit introduction of pion field. 
This aspect has been emphasized in the study of the Alternating-Layer-Spin 
structure in the condensed phase \cite{sup}.

\subsubsection{Charged Pion Condensation}

Next consider the charged pion condensation in neutron matter, which should 
have a direct relevance with neutron star phenomena.
Consider, e.g., the $\pi^+$ propagator:
\beq
D_{\pi^+}^{-1}(\omega, k;\rho_B)=\omega^2-m_\pi^2-k^2
{-\frac{\omega}{2f_\pi^2}\rho}-\Pi_p(\omega, k;\rho_B).
\eeq
Note that there appears the isovector $s$- wave coupling term 
$\propto (\rho_n-\rho_p)$ besides the $p$- wave term.
Poles of $D_{\pi^+}^{-1}$ include the energies of $\pi^\pm$ mesons and 
the $pn^{-1}$ {\it collective} mode with the same quantum 
number of $\pi^+$, called $\pi_s^+$, 
besides single ph and $\Delta$h excitations. The threshold condition for the 
charged pion condensation is 
\beq
\omega_{\pi^+_s}+\omega_{\pi^-}=0,
\eeq
which implies the $\pi_s^+\pi^- ~{\rm pair ~condensation}$.
In terms of the propagator we have 
\beq
D_{\pi^+}^{-1}(\omega=\mu_\pi^c, k_c,;\rho_c)=0, 
\left.\frac{\partial D_{\pi^+}^{-1}}{\partial k}\right|_{k=k_c}=0, 
\left.\frac{\partial D_{\pi^+}^{-1}}
{\partial \omega}\right|_{\omega=\mu_\pi^c}=0,
\eeq
which are called the {\it double-pole condition}.

The condensed phase can be represented as 
\beq
|\bpi^c\rangle=\exp(i\int V_3\bk_c\cdot\br d^3x)\exp(iQ_1^5\theta)
|{\rm normal}\rangle,
\eeq
and we see 
\beq
\langle\bpi\rangle=\bpi^c=(\sin\theta\cos\bk_c\cdot\br, 
\sin\theta\sin\bk_c\cdot\br, 0).
\eeq
Accordingly nucleons form the {quasi-particles $\eta,\zeta$} in the 
condensed state, which are 
superposition of {$p,n$}. Then the pion cooling works through the process;
\beq
\eta \rightarrow \eta'+e^-+\bar\nu_e
\eeq
and
\beq
\eta' +e^-\rightarrow \eta+\nu_e.
\eeq
It provide an efficient cooling mechanism for young neutron stars.

%\item {Potential description} may not be possible. 
%We need explicit {pion} degrees of freedom. 
%\end{itemize}

\subsection{Gamow-Teller Resonance and Critical Densities of \\
Pion Condensation}
\subsubsection{New Information on Landau-Migdal Parameters}

The Landau-Migdal parameters play a crucial role in the study of 
the spin-isospin excitation modes as well as the spin-
dependent structure of nuclei. Experimentally their values are not well known 
yet. Since experimental information was very limited, especially for 
$g'_{N\Delta}$ and $g'_{\Delta\Delta}$, one analyzed the spin dependent 
phenomena by using the {\it universality} ansatz,
\beq
g'_{NN}=g'_{N\Delta}=g'_{\Delta\Delta}{\equiv g'},
\eeq
e.g., based on the $SU(6)$ quark model \cite{mey}. Then it has been
found that many experimental data can be reproduced by the value of 
$g'=0.6 - 0.8$. Furthermore they concluded that 
pion condensation {\it does not occur} at several times nuclear density.

The giant Gamow-Teller (GT) states correspond to the spin-isospin dependent 
particle-hole states with small momentum, and thereby their coupling with 
pions should be small. Hence their excitation energy and strength 
should provide us with important 
information about the Landau-Migdal parameters. The strength, however, was 
not fully determined. At least about 50$\%$ of the GT sum-rule value was 
observed, but it was not clear why the rest of strength is missing \cite{ost}. 
Nevertheless, if one assumes that the strength is quenched owing to the 
coupling of the particle-hole states with the $\Delta$-hole states, the 
universality ansatz with $g'=0.6 - 0.9$ can explain well both the 
excitation energy and strength.    

Recent analysis of the giant GT resonance observed by
$
^{90}Zr (p,n) ^{90}Nb~ 
$
at $295$MeV suggests the quenching factor is small \cite{wak},
\beq
Q=\left[1-\frac{{ g'_{N\Delta}} U_\Delta^{(0)}}{1+
{g'_{\Delta\Delta}}U_\Delta^{(0)}}\right]=90\pm 5\%
\eeq
(c.f. Q=0.5 - 0.7 given by previous results).
Using also the excitation energy {$\omega_{GT}$}, we obtain \cite{suz}
\beqa
g'_{N\Delta}&=&0.18+0.05g'_{\Delta\Delta}\nonumber\\
g'_{NN}&\simeq& 0.59.
\eeqa
Thus two Landau-Migdal parameters have been experimentally fixed, while 
$g'_{\Delta\Delta}$ is still left as an unknown parameter.

\subsubsection{Critical Densities}

Using the new information about the Landau-Migdal parameters, we reexamine 
the possibility of pion condensations \cite{suz2}. In Figs~5, 6 the critical
densities of $\pi^0$ condensation are presented with those under the
universality ansatz for comparison. In the case of neutron matter we use 
$g_{NN}=g'_{NN}$ for simplicity, since the value of $g_{NN}$ is not
well-known yet. We can see that critical density moderately
increases as $g'_{\Delta\Delta}$ does, while the universality ansatz
gives a sharply increasing function; the critical densities result in 
low densities: $1<\rho_c/\rho_0<2.5$
for $g'_{\Delta\Delta}<1$. 

\begin{figure}[ht]
%\epsfxsize=10cm   %width of figure - will enlarge/reduce the figures
%\epsfbox{fig3.eps}
%\figurebox{2cm}{3cm}{} %to have a box alone 
\centerline{\epsfxsize=4.5in\epsfbox{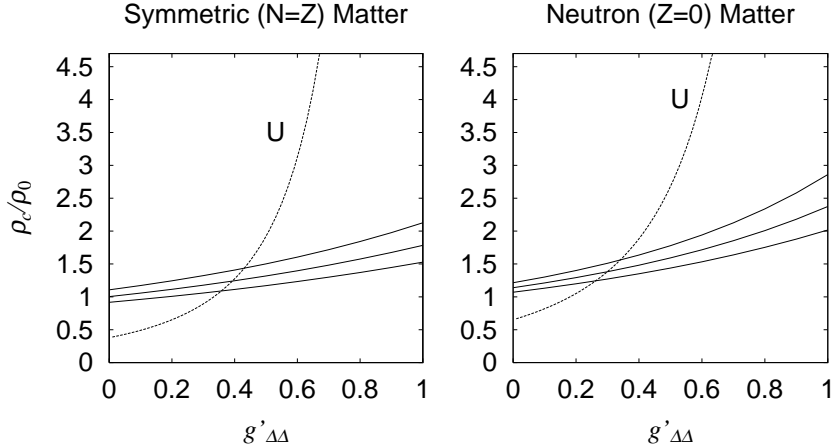}}   
\caption{Critical densities for neutral pion condensation in symmetric 
nuclear matter (left panel) and pure neutron matter (right panel). Three 
solid lines in the left panel show those in the three effective-mass 
cases from 
$m^*_N=0.8m_N$ (bottom) to $m^*_N=0.6m_N$ (top) by $0.1m_N$ step. 
Similarly, the 
solid lines in the right panel show those in the three $Q$ cases from 
$Q=0.85$ (top) to $Q=0.95$ (bottom) by $0.05$ step. The dashed curve denoted 
by ``U'' means those by the universality ansatz.}
\end{figure}

Recently sophisticated variational calculations have been done for
symmetric nuclear matter and pure neutron matter, using modern 
potentials \cite{akm,hei}. They also found that there are phase transitions
to pion condensation at low densities, $2\rho_0$ and $1.3\rho_0$ for
symmetric nuclear matter and pure neutron matter, respectively. It would
be interesting to compare these values with our results.

The critical density for charged pion condensation is presented in
Fig.~6. The behavior is almost the same as that for $\pi^0$ condensation 
and the critical density is low; $\rho_0<\rho_c<2\rho_0$ for 
$g'_{\Delta\Delta}<1$ and $m^*_N=0.8m_N$.
It would be interesting to refer the works by 
Tsuruta et al. \cite{tsu,ume} in this
context: they set charged pion condensation at $\rho_c=2.5\rho_0$ in 
their calculation of neutron star cooling.

\begin{figure}[ht]
%\epsfxsize=10cm   %width of figure - will enlarge/reduce the figures
%\epsfbox{fig3.eps}
%\figurebox{2cm}{3cm}{} %to have a box alone 
\centerline{\epsfxsize=2.5in\epsfbox{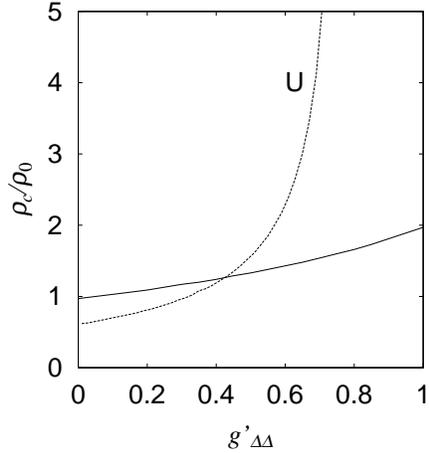}}   
\caption{Critical density for charged pion condensation in pure neutron 
matter. $m^*_N=0.8m_N$ and $Q=0.9$ are used here. The symbol ``U'' means
 that by the universality ansatz.}
\end{figure}

%pandha.
%ALS

\section{Summary and Concluding Remarks}

We have seen three recent results, which may support possible existence of 
pion condensation at low densities. The new experiment on the Gamow-Teller 
resonance tells us the universality ansatz about the Landau-Migdal 
parameters by no means hold; $g'_{N\Delta}$ should be much less than 
$g'_{NN}$ or $g'_{\Delta\Delta}$. The critical densities of pion condensations 
are $\rho_c\sim 1.5\rho_0(N=Z, 
m^*_N/m_N=0.8, g'_{\Delta\Delta}=1)$ and 
$\rho_c\sim 2.4\rho_0(Z=0, m^*_N/m_N=0.8, g'_{\Delta\Delta}=1)$ for 
neutral pion 
condensation, while 
$\rho_c\sim 2.\rho_0(Z=0, m^*_N/m_N=0.8, g'_{\Delta\Delta}=1)$ for 
charged pion condensation.

As another theoretical work, a new calculation  of 
nuclear matter with a modern potential has also suggested the phase 
transition at $\rho_c\sim 2\rho_0(N=Z)$ and $\rho_c\sim 1.3\rho_0(Z=0)$, 
which corresponds to neutral pion condensation \cite{akm}.

Besides these theoretical developments, the current observation about the 
surface temperature of a neutron star inside 3C58 suggests we need 
exotic cooling mechanisms beyond the standard cooling scenario. We have seen 
that a consistent calculation about pion cooling have been done by taking 
into account nucleon superfluidity in a proper way, and it can explain 
the data \cite{tsu}. 

Unfortunately these are indirect evidences for pion condensation, and we hope 
 for a direct evidence by heavy-ion collision experiments in near future.

Finally I would like give a comment about another theoretical aspect 
of pion condensation. It means a spontaneous violation of parity 
in nuclear matter and the condensed phase can be described as 
a chirally rotated state. We may also consider its analog in quark matter:
nonvanishing of the parity-odd mean-filed 
$\langle\bar q\gamma_5\Gamma^\alpha q\rangle$. It would be interesting 
in this context to 
refer to recent studies about ferromagnetism in quark matter, where a 
magnetic ordering is realized under the axial-vector mean-field \cite{tat}.

\section*{Acknowledgments}

The author thanks T. Suzuki, H. Sakai, M. Nakano, S. Tsuruta,
T. Takatsuka, T. Muto and 
R. Tamagaki  for their collaboration.

\end{document}